\documentclass[prl,twocolumn,showpacs,preprintnumbers,amsmath,amssymb]{revtex4}

\usepackage{graphicx}
\usepackage{dcolumn}
\usepackage{bm}
\usepackage{subfigure}

\newcommand{\ket}[1]{|#1\rangle}

\begin{document}

\title{Spectroscopy of atomic rubidium at 500 bar buffer gas pressure:
approaching the thermal equilibrium of dressed atom-light states}

\author{Ulrich Vogl}
\email{vogulr@iap.uni-bonn.de}
\author{Martin Weitz}%
\affiliation{
Institut f\"ur Angewandte Physik der Universit\"at Bonn, Wegelerstra\ss e 8,
 53115 Bonn, Germany,}
 \affiliation{
Physikalisches~\hspace{-0.3mm}Institut~\hspace{-0.3mm}der~\hspace{-0.3mm}Universit\"at~\hspace{-0.3mm}T\"ubingen,~\hspace{-0.3mm}Auf~\hspace{-0.3mm}der~\hspace{-0.3mm}Morgenstelle~\hspace{-0.3mm}14,~\hspace{-0.3mm}72076~\hspace{-0.3mm}T\"ubingen,~\hspace{-0.3mm}Germany}


\date{\today}

\begin{abstract}
We have recorded fluorescence spectra of the atomic rubidium
D-lines in the presence of several hundreds of bars buffer gas
pressure. With additional saturation broadening a spectral
linewidth comparable to the thermal energy of the atoms in the
heated gas cell is achieved.  An intensity-dependent blue
asymmetry of the spectra is observed, which becomes increasingly
pronounced when extrapolating to infinitely high light intensity.
We interpret our results as evidence for the dressed (coupled
atom-light) states to approach thermal equilibrium.
\end{abstract}

\maketitle

In usual optical spectroscopy experiments, light sources along with
spontaneous decay drive the investigated sample far from thermal
equilibrium \cite{demt}. This is most evident in room temperature atomic physics
experiments, where the observed spectral linewidths in energy units
are many orders of magnitude
below the thermal energy $k_BT$ ($\simeq 1/40$ eV for $T=$ 300 K). Thus, apart
from a comparatively small and symmetric broadening arising from the known
Doppler effect, the excitation profile is largely independent of the
statistical distribution function. In cold atom experiments, spectral
linewidth and thermal energy can be comparable. However, the lack of a
sufficiently fast thermalization process has so far here prevented thermal
equilibrium of coupled atom-light states to
be a useful concept.

We here report on experiments, in which up to 500 bar of argon and
helium respectively buffer gas pressure induce a linewidth of a
few nanometers for the rubidium D-lines. At high optical power of
the exciting continuous-wave laser source, the fluorescence
spectra are broadened by additional power broadening to values
exceeding the thermal energy $k_{B}T$ in the heated gas cell. In
this regime, we observe a strong blue asymmetry of the spectra.
The spectral asymmetry increases further when extrapolating our
data towards infinitely high excitation intensity. We interpret
our results as evidence for the coupled atom-light states
(''dressed states'') to approach thermal equilibrium, with the
thermalization being due to frequent rubidium-buffer gas
collisions. Thermal equilibrium is e.g.\ a prerequisite for
possible BEC-like phase transitions of coupled atom-light
quasiparticles (polaritons) \cite{eal}.

Our experiment benefits of collisional aided excitation, which
allows to saturate the atomic transition at large detuning with a
continuous-wave laser source. Pressure broadening of atomic
spectral lines is a long investigated topic
\cite{RevModPhys.29.20, 1974PhR....12..273S, Royer, CiuryloSzudy,
alioua:032711, PhysRevA.51.1085, PhysRevA.53.1183,Stienke}. For
large laser detunings ($\gg 1/\tau_{coll}$, where $\tau_{coll}$
denotes the collisonal duration), the impact limit ceases to be
fulfilled, and atomic lines often become asymmetric. A theoretical
description of the absorption probability requires knowledge of
the molecular potential curves of the collisional partners
\cite{PhysRevA.6.1519}, and can be expressed e.g.~in terms of
modulated dipole approaches \cite{PhysRevA.19.1106}. The effect of
collisional redistribution has shown the influence of state
changing collisions \cite{Liao}. In an interesting experiment,
deviations from the Einstein coefficients for absorption and
stimulated emission have been observed for far off-resonant
excitation in a collisionally broadened system\cite{markov}.
Dressed state approaches have proven to be an elegant way to treat
collisional processes in the presence of laser radiation
\cite{ct}.

For a theoretical description of the transition to thermal
equilibrium of dressed states, we give a simple model based on
thermodynamic arguments. Consider a two-level system with ground
state $\ket{g}$ and excited state $\ket{e}$ coupled to a laser
field of frequency $\omega$. When tuned into resonance, the laser
field connects the states $\ket{g, n+1}$ and $\ket{e,n}$
respectively, where the first quantum number denotes the internal
atomic state and the second one the photon number. To simplify the
analysis, let us restrict ourselves to the case of large detunings
$\delta$~=~$\omega$~-~$\omega_{atom}$, with $\left|\delta\right|
\gg \Omega _{Rabi}$, so that in the absence of collisions both the
state mixing and the AC Stark shift can be neglected. The energy
splitting between dressed states is thus simply $\delta$, as shown
in Fig.~1a. On the other hand, when a buffer gas atom approaches
the energy levels are shifted, and efficient excitation and a
transfer between dressed state energy levels is provided when the
laser field transiently becomes resonant during the collision,
yielding an example of collisional aided excitation. Let
$c_{g\rightarrow e}(\delta )$ and $c_{e\rightarrow g}(\delta )$
denote the rate constants for collisional transfer between the
dressed state levels, which for $\left|\delta\right| \gg \Omega
_{Rabi}$ can be shown to be proportional to the laser intensity
\cite{ct}. For a finite spontaneous decay rate $\Gamma$, also a
coupling between states $\ket{e,n}$ and $\ket{g,n}$ of different
dressed state manifolds occurs.

We are interested in the properties of such a two level system
subject to laser radiation and coupled to a thermal bath. If we
assume our system to be optically thin, we can trace over the
photonic quantum numbers, and obtain the following Boltzmann-like
rate equations
\begin{figure}[t!!]
    \centering
     \includegraphics[width=8.6cm]{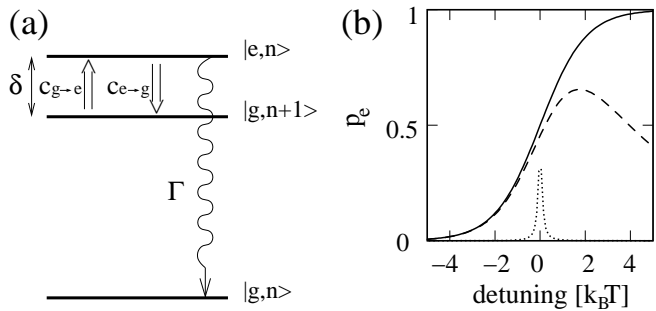}
     \caption{\label{fig:twolevel}(a) Scheme of levels and rates for a two-level dressed
state system approaching thermal equilibrium in the far detuned limit.
(b) Upper state population
as a function of laser frequency, where a Lorentzian lineshape
was taken for the detuning
dependent excitation rate $c(\delta)$: For $c(k_BT)\ll \Gamma$, i.e.~small
linewidth, which yields a lineshape resembling that of $c(\delta)$ (dotted
line), $c(k_BT)\simeq \Gamma$ (dashed
line) and $c(\delta )\gg \Gamma$, i.e.~thermal equilibrium of dressed
states (solid line).}
\vspace{-4mm}
\end{figure}
\begin{subequations}
\vspace{-1mm}
\begin{equation}
\label{1a}\dot{f_{e}}=c_{g\rightarrow e}(\delta )f_{g}-c_{e\rightarrow g}(\delta )f_{e}-\Gamma f_{e}
\end{equation}
\vspace{-9mm}
\begin{equation}
\label{1b}\dot{f_{g}}=c_{e\rightarrow g}(\delta )f_{e}-c_{g\rightarrow e}(\delta )f_{g}+\Gamma f_{e},
\vspace{-1mm}
\end{equation}
\end{subequations}
where $f_{e}$ and $f_{g}$ are the
corresponding populations of states $\ket{e}$ and $\ket{g}$.
To derive a relation between the rates $c_{g\rightarrow e}$ and $c_{e\rightarrow
g}$, let us consider the equilibrium solution ($\dot{f_{e}}=\dot{f_{g}}=0$)
in the absence of spontaneous decay (i.e.~$\Gamma\rightarrow 0$). In this
limit, we expect that usual Boltzmann statistics can be applied to
this two-level system, which yields $f_{e}=e^{\delta/k_BT}/(1+e^{\delta/k_BT})=1/(1+e^{-\delta/k_BT})$, which resembles
the Fermi-Dirac distribution function due to only two available
states. From Eqs.~1 we obtain $f_{e}=1-f_{g}=1/(1+c_{e\rightarrow g}/c_{g\rightarrow e})$
so that the following relation of the rate
constants must apply: $c_{e\rightarrow g}/c_{g\rightarrow e}=e^{-\delta/k_BT}$, as also shown in \cite{markov}.
Note that the sign of the laser detuning in this formula depends on the
definition of $\delta$.
Physically, the increased rate towards a population of the energetically
lower dressed state is due to the larger phase space that becomes available to
the reservoir with the associated energy gain of $\left|\delta\right|$. The result is also
obtained when considering the variation of the available density
of states in Fermi's golden rule.

Spontaneous decay drives the dressed state system out of thermal
equilibrium. The full Eq.~1 with $\dot{f_{e}}=\dot{f_{g}}=0$
yields the following stationary solution in our model:
\vspace{-1mm}
\begin{equation}
f_{e}=\frac{c(\delta )/\Gamma }{1+\frac{c(\delta )}{\Gamma }\cdot
(1+e^{-\delta /kT})}
\label{eq:1}
\vspace{-1mm}
\end{equation}
The condition for this formula to reduce to the usual Fermi-Dirac
distribution of a two-level system at large transition linewidth
and laser intensity is that $c(\delta )\gg \Gamma$ is fulfilled at
all relevant detunings $\delta$. On the other hand, for
$k_BT\gg\left|\delta\right|$ we obtain $f_{e}\simeq (c(\delta
)/\Gamma)/ (1+2c(\delta )/\Gamma )$, which corresponds to the
usual saturation dependence of a two-level system \cite{ct}.
Fig.~1b shows a comparison of line profiles in the different
regimes, where a Lorentzian excitation profile was assumed for the
detuning dependent excitation rate $c(\delta)$. The dotted line
gives the spectrum in the limit of a linewidth of $c(\delta)$
considerably below $k_BT$, resulting in a lineshape resembling
that of $c(\delta)$, while the solid line gives the case of a
linewidth considerably above the thermal energy (solid line).
Also, an example for the intermediate regime of comparable
linewidth and thermal energy is given (dashed line). A theoretical
prediction of the detuning dependent excitation rate $c(\delta)$
in the here investigated dense buffer gas regime is a highly
nontrivial task, especially since we expect multiple particle
collisions to play an important role. On the other hand, if full
thermal equilibium can be achieved for all used laser detunings,
the predicted lineshape reduces to a Fermi-Dirac profile
independently of the form of $c(\delta)$.

\begin{figure}[ht!!]
\includegraphics[width=8.6cm]{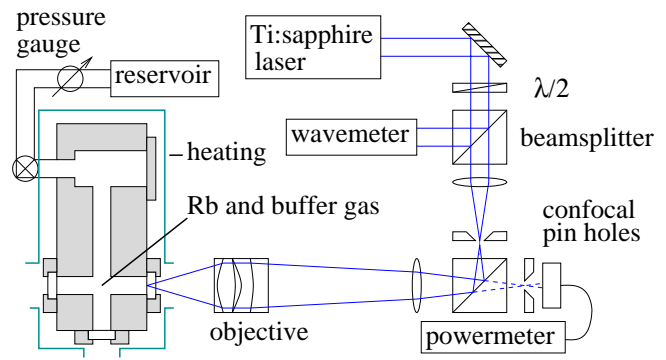}
\caption{\label{fig:set} Scheme of the experimental setup.}
\vspace{-0.5mm}
\end{figure}

A scheme of our experimental setup used to investigate rubidium
atoms at high buffer gas pressures and intense laser radiation is
shown in Fig.~2. A rubidium filled stainless steel oven with
volume of a few cm$^{3}$ and wall thickness of roughly a cm is
used as a pressure cell. Optical radiation is coupled into the
chamber through sapphire windows. Buffer gas is filled into the
chamber through an attached valve with pressures up to 230 bar (at
room temperature). Due to technical reasons, the pressure is
measured outside the chamber. During operation, the cell is heated
to 260~$^{\circ}$C,
which results in a vapour pressure limited rubidium number density of $%
\simeq~1.0\cdot 10^{16}/$cm$^{3}$ \cite{nes}. In the sealed chamber, the buffer
gas pressure then increases up to approximately 400 bar for helium and
500 bar for argon.

Tunable laser radiation for excitation of the atoms is provided by a
Ti:sapphire ring laser. Laser frequency scanning
over the broad atomic absorption profile is done manually within the
operation range of the used laser mirror set of
350-395 THz. The laser frequency is measured with a
wavemeter. With a Fraunhofer achromat, the laser radiation
is focused into the rubidium chamber to a beam radius of $\sim$ 3 $\mu$m,
which at a maximum optical power of 300 mW results in an intensity of 1.1$%
\cdot 10^{8}$ mW/cm$^{2}$. The beam focal plane within the cell is placed near
the sapphire window. To selectively record the atomic fluorescence only from
regions of high laser intensity, we furthermore spatially filter both the
incident optical beam and the outgoing fluorescence with pinholes in a
confocal geometry. The recorded fluorescence power values exhibited short-term
variations that became increasingly severe for a focal spot very close to
the cell window, which attribute to thermal fluctuations. For our
experimental spectra, the mean fluorescence power value averaged over 30
seconds is taken for every data point.

In initial experiments, we have measured the lifetime of the
rubidium 5P state at high argon buffer gas pressures. For this
measurement, an acoustooptic modulator (not shown in Fig.~2) was
included in the beam path, which allowed for a rapid shutting of
the exciting optical radiation for up to $\sim 30$ mW beam power.
The subsequent decay of the atomic fluorescence was monitored with
a photomultiplier tube. At low buffer gas pressures ($\lesssim 10$
bar) we observed fluorescence decay times up to 1 $\mu$s, which we
attribute to repopulation from higher states due to energy pooling
\cite{ekers}. More direct evidence for energy pooling is obtained
from the clearly visible blue fluorescence at those pressure
values. At higher buffer gas pressures the fluorescence signal
decayed with a time constant of 40 ns. If we account for the time constants of our acoustooptic modulator and the photomultiplier detection, this value is in
good agreement with the lifetime value of about 27 ns obtained in
earlier measurements \cite{rb}. We conclude that energy pooling to
higher state excitation is suppressed at sufficient buffer gas
pressure. Further, the lifetime of the 5P$_{1/2}$ and 5P$_{3/2}$
states are, within our measurement accuracy of 5 ns, unchanged to
argon pressures of 200 bar, which is the maximum achieved
pressure value in these early data sets. At a collisional rate of 10$%
^{11}/$s, the atoms experience more than 10$^{3}$ collisions
within a natural lifetime. This confirms the remarkable elasticity
of excited atoms with rare gas collisions \cite{noquench}, and is
a prerequisite for the thermalization of dressed states in our
experimental scheme. Note that the excited state lifetime here is
some four orders of magnitude above the $\sim$ps lifetimes
observed in semiconductor exciton systems \cite{eal}.

\begin{figure}[t!!]
       \centering
       \includegraphics[width=8.6cm]{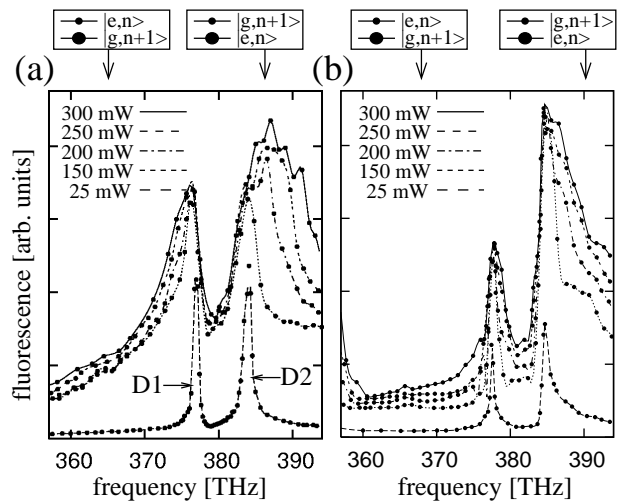}
       \caption{\label{fig:all} Fluorescence spectrum of the rubidium D-lines at 500 bar argon (a) and 400 bar helium (b) buffer gas
pressure for different optical beam powers. The small drawings on the top indicate the population
of the dressed states on the red and blue side of the electronic transition
respectively.}
\vspace{-1mm}
 \end{figure}
Typical fluorescence spectra recorded under high pressure buffer
gas conditions for variable light intensities are shown in Fig.~3a
(argon buffer gas) and Fig.~3b (helium buffer gas). While the observed lineshapes somewhat differ for both used buffer
gases, indicating the 'transiently chemical' nature of optical
collisions in a regime beyond the impact limit, let us draw our
attention to some general features for both spectra. At moderate
optical power ($P\simeq 25$ mW) the linewidths are clearly below
the thermal energy $k_BT$ ($\simeq 11$ THz at $T=530$ K). At
higher power levels the linewidths increase to values around
$k_BT$. Noticeably, the red side of the D1-line tends to saturate
to a comparatively low fluorescence level, while at the blue side
of the D2-line much higher fluorescence is observed, though in the
far wings even at maximum power levels (300 mW) the transition is
less saturated. We attribute this to the dressed state system
approaching thermal equilibrium for high power levels. The
resonant Rabi coupling at full optical power is of order 100 GHz,
so that for detunings of order $k_{B}T$  the mixing of ground and
excited states is still small. For the sake of simplicity, let us
restrict the discussion to the red side of the D1-line and the
blue side of the D2-line, as in these limits the influence of the
upper state fine structure splitting is smallest. While on the red
side of a two-level system the state $\ket{g,n+1}$ is
energetically below $\ket{e,n}$, leading to an enhanced ground
state population in thermal equilibrium, on the blue side of the
transition the dressed excited state component is lowest and
expected to be favored in equilibrium, as indicated in the boxes
in the top of Fig.~3. At low optical power levels, the upper state
spontaneous decay drives the coupled atom light system away from
equilibrium.
\begin{figure}[t!!]
    \centering
      \includegraphics[width=8.6cm]{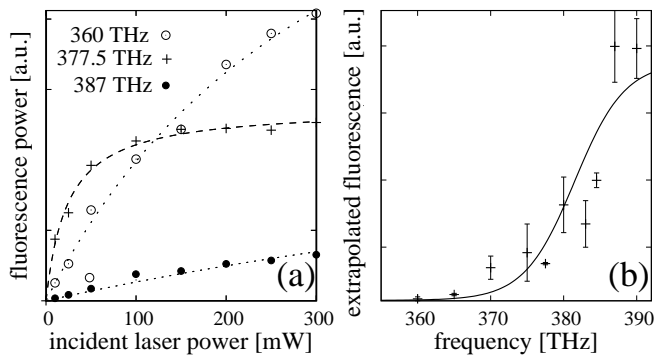}
      \caption{\label{fig:sat}(a) fluorescence of rubidium atoms at
      400 bar helium buffer gas as a function of optical power for
different laser frequencies. The experimental data has been fitted with the
expected saturation dependence of Eq.~(\ref{eq:1}). (b) Expected atomic fluorescence
at infinite laser power, obtained from extrapolation curves as in (a),
versus laser frequency. The shown error bars were obtained
from the extrapolation fits. The data has been fitted with a
Fermi-Dirac distribution, assuming thermal equilibrium of the two-level
dressed state system.}
\end{figure}
For a more detailed analysis, Fig.~4a shows the observed
fluorescence as a function of the optical power for three
different laser frequencies for the data with helium buffer gas.
These saturation curves have been fitted with the theoretical
prediction of Eq.~(\ref{eq:1}), where a constant prefactor and the
ratio of $c(\delta)$ and the laser intensity were left as free fit
parameters. This fitting procedure assumes that the detuning
dependent excitation rate $c(\delta)$ is linearly dependent on the
laser intensity, an assumption in agreement with present theories
of optical collisions, whose validity in the yet largely unexplored
high pressure buffer gas regime with large multiparticle
collisional rate however remains to be tested \cite{pers_comm}.
Near a line center (crosses), the curve saturates already at
relatively low power level to an intermediate fluorescence level.
For significant detuning the optical power at which saturation is
achieved is higher, as visible both for blue (dots) and red
(circles) detunings. Interestingly, the saturation level of
fluorescence is clearly different in both cases, with the blue
(red) detuning leading to largest (smallest) values.

Fig.~4b shows the value of fluorescence extrapolated to infinite
beam power, as taken from such curves, as a function of laser
frequency. This is to provide an estimate for the upper state
population at thermal equilibrium. The data has been fitted with a
Fermi-Dirac distribution $f _{e}(\delta )=1/(1+e^{-\delta /k_BT})$
\ expected for the upper state population of the two level dressed
state system, where the detuning $\delta$ was measured relatively
to the center of the D-lines. The agreement between theory
and experiment is quite reasonable, which supports the assumption
of the dressed states to approach thermal equilibrium at high
laser power. An issue remaining to be clarified is that a best fit
is obtained with a temperature $T=162(58)$K, which is
significantly below the cell temperature. For a comparable fit
with the argon buffer gas spectral data, a temperature of
$270(91)$K was obtained. At present, the origin of this
disagreement of temperatures is not resolved. A more detailed rate
equation model of partial thermal equilibrium should include the
rubidium fine structure and spatially inhomogeneous saturation. On
the other hand, an interesting question is whether laser cooling
(heating) processes of the thermal gas here occur on the red
(blue) side of the transition due to the energy loss (gain) during
collisional aided excitation, as was proposed in
\cite{Berman_cool}. If the spontaneously emitted photons are
energy shifted by an amount comparable to $k_{B}T$ respectively to
the incident photon wavelength, cooling may be possible in such
pressure broadened systems, comparable to results obtained on the
optical cooling of solids \cite{PhysRevLett.78.1030}.

To conclude, we have recorded spectra of rubidium atoms in a collisionally
broadened regime interpolating between usual atomic physics gas phase and
solid/liquid phase conditions. With additional saturation broadening, the
spectral linewidth approaches the thermal energy. The saturation dependence of the
line profile is interpreted as evidence for the
onset of thermal equilibrium of dressed atom-light states.

For the future, it would be interesting to study possible novel laser cooling
mechanisms of high pressure atomic gases based on collisionally aided
excitation. A quite
different perspective could include a possible BEC-like phase transition to
a condensed atom-light polariton phase. For our present apparatus, this
would require an increase of the optical intensity and an optically thick
atomic ensemble, as could be realized with a cavity- or optical waveguide
based system.

We acknowledge helpful discussions with N. Schopohl and R. Walser. This work
has been funded within a collaborative transregional research center (TR 21)
of the DFG.

\begin{thebibliography}{21}
\expandafter\ifx\csname natexlab\endcsname\relax\def\natexlab#1{#1}\fi
\expandafter\ifx\csname bibnamefont\endcsname\relax
  \def\bibnamefont#1{#1}\fi
\expandafter\ifx\csname bibfnamefont\endcsname\relax
  \def\bibfnamefont#1{#1}\fi
\expandafter\ifx\csname citenamefont\endcsname\relax
  \def\citenamefont#1{#1}\fi
\expandafter\ifx\csname url\endcsname\relax
  \def\url#1{\texttt{#1}}\fi
\expandafter\ifx\csname urlprefix\endcsname\relax\def\urlprefix{URL }\fi
\providecommand{\bibinfo}[2]{#2}
\providecommand{\eprint}[2][]{\url{#2}}

\bibitem[{\citenamefont{Demtr{\"o}der}(2003)}]{demt}
\bibinfo{author}{\bibfnamefont{See, e.g.: W.}~\bibnamefont{Demtr{\"o}der}},
  \emph{\bibinfo{title}{Laser Spectroscopy}} (\bibinfo{publisher}{Springer,
  Berlin}, \bibinfo{year}{2003}).

\bibitem[{\citenamefont{Eastham and Littlewood}(2001)}]{eal}
\bibinfo{author}{\bibfnamefont{P.~R.} \bibnamefont{Eastham}} \bibnamefont{and}
  \bibinfo{author}{\bibfnamefont{P.~B.} \bibnamefont{Littlewood}},
  \bibinfo{journal}{Phys.\ Rev. B} \textbf{\bibinfo{volume}{47}},
  \bibinfo{pages}{235101} (\bibinfo{year}{2001}).

\bibitem[{\citenamefont{Ch'en and Takeo}(1957)}]{RevModPhys.29.20}
\bibinfo{author}{\bibfnamefont{S.-y.} \bibnamefont{Ch'en}} \bibnamefont{and}
  \bibinfo{author}{\bibfnamefont{M.}~\bibnamefont{Takeo}},
  \bibinfo{journal}{Rev. Mod. Phys.} \textbf{\bibinfo{volume}{29}},
  \bibinfo{pages}{20} (\bibinfo{year}{1957}).

\bibitem[{\citenamefont{{Schuller} and
  {Behmenburg}}(1974)}]{1974PhR....12..273S}
\bibinfo{author}{\bibfnamefont{F.}~\bibnamefont{{Schuller}}} \bibnamefont{and}
  \bibinfo{author}{\bibfnamefont{W.}~\bibnamefont{{Behmenburg}}},
  \bibinfo{journal}{Phys. Rep.} \textbf{\bibinfo{volume}{12}},
  \bibinfo{pages}{273} (\bibinfo{year}{1974}) (\bibinfo{author}{and references therein}).

\bibitem[{\citenamefont{Royer}(1980)}]{Royer}
\bibinfo{author}{\bibfnamefont{A.}~\bibnamefont{Royer}},
  \bibinfo{journal}{Phys. Rev. A} \textbf{\bibinfo{volume}{22}},
  \bibinfo{pages}{1625} (\bibinfo{year}{1980}).

\bibitem[{\citenamefont{Ciury\l{}o and Szudy}(2001)}]{CiuryloSzudy}
\bibinfo{author}{\bibfnamefont{R.}~\bibnamefont{Ciury\l{}o}} \bibnamefont{and}
  \bibinfo{author}{\bibfnamefont{J.}~\bibnamefont{Szudy}},
  \bibinfo{journal}{Phys. Rev. A} \textbf{\bibinfo{volume}{63}},
  \bibinfo{pages}{042714} (\bibinfo{year}{2001}).

\bibitem[{\citenamefont{Stienkemeier et~al.}(1996)\citenamefont{Stienkemeier,
  Higgins, Callegari, Kanorsky, Ernst, and Scoles}}]{Stienke}
\bibinfo{author}{\bibfnamefont{F.}~\bibnamefont{Stienkemeier, et al.}},
    \bibinfo{journal}{Z. Phys. D} \textbf{\bibinfo{volume}{38}},
  \bibinfo{pages}{253} (\bibinfo{year}{1996}).

\bibitem[{\citenamefont{Alioua and Bouledroua}(2006)}]{alioua:032711}
\bibinfo{author}{\bibfnamefont{K.}~\bibnamefont{Alioua}} \bibnamefont{and}
  \bibinfo{author}{\bibfnamefont{M.}~\bibnamefont{Bouledroua}},
  \bibinfo{journal}{Phys. Rev. A} \textbf{\bibinfo{volume}{74}},
  \bibinfo{eid}{032711} (\bibinfo{year}{2006}).

\bibitem[{\citenamefont{Kristensen et~al.}(1995)\citenamefont{Kristensen, Blok,
  van Eijkelenborg, Nienhuis, and Woerdman}}]{PhysRevA.51.1085}
\bibinfo{author}{\bibfnamefont{M.}~\bibnamefont{Kristensen, et al.}},
    \bibinfo{journal}{Phys. Rev. A} \textbf{\bibinfo{volume}{51}},
  \bibinfo{pages}{1085} (\bibinfo{year}{1995}).

\bibitem[{\citenamefont{Woerdman et~al.}(1996)\citenamefont{Woerdman, Blok,
  Kristensen, and Schrama}}]{PhysRevA.53.1183}
\bibinfo{author}{\bibfnamefont{J.~P.} \bibnamefont{Woerdman, et al.}},
   \bibinfo{journal}{Phys. Rev. A}
  \textbf{\bibinfo{volume}{53}}, \bibinfo{pages}{1183} (\bibinfo{year}{1996}).

\bibitem[{\citenamefont{Hedges et~al.}(1972)\citenamefont{Hedges, Drummond, and
  Gallagher}}]{PhysRevA.6.1519}
\bibinfo{author}{\bibfnamefont{R.~E.~M.} \bibnamefont{Hedges}},
  \bibinfo{author}{\bibfnamefont{D.~L.} \bibnamefont{Drummond}},
  \bibnamefont{and}
  \bibinfo{author}{\bibfnamefont{A.}~\bibnamefont{Gallagher}},
  \bibinfo{journal}{Phys. Rev. A} \textbf{\bibinfo{volume}{6}},
  \bibinfo{pages}{1519} (\bibinfo{year}{1972}).

\bibitem[{\citenamefont{Yeh and Berman}(1979)}]{PhysRevA.19.1106}
\bibinfo{author}{\bibfnamefont{S.}~\bibnamefont{Yeh}} \bibnamefont{and}
  \bibinfo{author}{\bibfnamefont{P.~R.} \bibnamefont{Berman}},
  \bibinfo{journal}{Phys. Rev. A} \textbf{\bibinfo{volume}{19}},
  \bibinfo{pages}{1106} (\bibinfo{year}{1979}).

\bibitem[{\citenamefont{Liao et~al.}(1979)\citenamefont{Liao, Bjorkholm, and
  Berman}}]{Liao}
\bibinfo{author}{\bibfnamefont{P.~F.} \bibnamefont{Liao}},
  \bibinfo{author}{\bibfnamefont{J.~E.} \bibnamefont{Bjorkholm}},
  \bibnamefont{and} \bibinfo{author}{\bibfnamefont{P.~R.}
  \bibnamefont{Berman}}, \bibinfo{journal}{Phys. Rev. A}
  \textbf{\bibinfo{volume}{20}}, \bibinfo{pages}{1489} (\bibinfo{year}{1979}).

\bibitem[{\citenamefont{Markov et~al.}(2002)\citenamefont{Markov, Plekhanov,
  and Shalagin}}]{markov}
\bibinfo{author}{\bibfnamefont{R.~V.} \bibnamefont{Markov}},
  \bibinfo{author}{\bibfnamefont{A.~I.} \bibnamefont{Plekhanov}},
  \bibnamefont{and} \bibinfo{author}{\bibfnamefont{A.~M.}
  \bibnamefont{Shalagin}}, \bibinfo{journal}{Phys. Rev. Lett.}
  \textbf{\bibinfo{volume}{88}}, \bibinfo{pages}{213601}
  (\bibinfo{year}{2002}).

\bibitem[{\citenamefont{Cohen-Tannoudji
  et~al.}(1992)\citenamefont{Cohen-Tannoudji, Dupont-Roc, and Grynberg}}]{ct}
\bibinfo{author}{\bibfnamefont{C.}~\bibnamefont{Cohen-Tannoudji}},
  \bibinfo{author}{\bibfnamefont{J.}~\bibnamefont{Dupont-Roc}},
  \bibnamefont{and} \bibinfo{author}{\bibfnamefont{G.}~\bibnamefont{Grynberg}},
  \emph{\bibinfo{title}{Atom-Photon Interactions
}} (\bibinfo{publisher}{Wiley, New York}, \bibinfo{year}{1992}).

\bibitem[{\citenamefont{Nesmeyanow}(1963)}]{nes}
\bibinfo{author}{\bibfnamefont{A.~N.} \bibnamefont{Nesmeyanow}},
  \emph{\bibinfo{title}{Vapor Pressure of the Chemical Elements}}
  (\bibinfo{publisher}{Elsevier, Amsterdam}, \bibinfo{year}{1963}).

\bibitem[{\citenamefont{Ekers and et~al.}(2001)}]{ekers}
\bibinfo{author}{\bibfnamefont{A.}~\bibnamefont{Ekers, et~al.}}, \bibinfo{journal}{Can. J. Phys.
  {\bf{79}}, 1039}  (\bibinfo{year}{2001}).

\bibitem[{\citenamefont{{Volz} and {Schmoranzer}}(1996)}]{rb}
\bibinfo{author}{\bibfnamefont{U.}~\bibnamefont{{Volz}}} \bibnamefont{and}
  \bibinfo{author}{\bibfnamefont{H.}~\bibnamefont{{Schmoranzer}}},
  \bibinfo{journal}{Phys. Scr. T} \textbf{\bibinfo{volume}{65}},
  \bibinfo{pages}{48} (\bibinfo{year}{1996}).

\bibitem[{\citenamefont{Speller et~al.}(1979)\citenamefont{Speller,
  Staudenmayer, and Kempter}}]{noquench}
\bibinfo{author}{\bibfnamefont{E.}~\bibnamefont{Speller}},
  \bibinfo{author}{\bibfnamefont{B.}~\bibnamefont{Staudenmayer}},
  \bibnamefont{and} \bibinfo{author}{\bibfnamefont{V.}~\bibnamefont{Kempter}},
  \bibinfo{journal}{Z. Phys. A} \textbf{\bibinfo{volume}{291}},
  \bibinfo{pages}{311} (\bibinfo{year}{1979}).

\bibitem[{\citenamefont{{Berman} and {Stenholm}}(1978)}]{Berman_cool}
\bibinfo{author}{\bibfnamefont{P.~R.} \bibnamefont{{Berman}}} \bibnamefont{and}
  \bibinfo{author}{\bibfnamefont{S.}~\bibnamefont{{Stenholm}}},
  \bibinfo{journal}{Opt. Commun.} \textbf{\bibinfo{volume}{24}},
  \bibinfo{pages}{155} (\bibinfo{year}{1978}).

\bibitem{pers_comm}
\bibnamefont{G. Pichler, private communications}.


\bibitem[{\citenamefont{Mungan et~al.}(1997)\citenamefont{Mungan, E., Buchwald,
  Edwards, Epstein, and Gosnell}}]{PhysRevLett.78.1030}
\bibinfo{author}{\bibfnamefont{C. E.}~\bibnamefont{Mungan, et al.}},
   \bibinfo{journal}{Phys. Rev. Lett.}
  \textbf{\bibinfo{volume}{78}}, \bibinfo{pages}{1030} (\bibinfo{year}{1997}).


\end{thebibliography}

\end{document}